\documentclass[twocolumn,prb,showkeys,superscriptaddress,showpacs,preprintnumbers,floatfix]{revtex4}
\usepackage{amssymb}
\usepackage{graphicx}

\begin{document}

\title{Multichannel field-effect spin barrier selector}
\author{G. E. Marques}
\affiliation{Departamento de F\'{\i}sica, Universidade Federal de S\~{a}o Carlos,
13.565-905, S\~{a}o Carlos, S\~{a}o Paulo, Brazil}
\author{A. C. R. Bittencourt}
\affiliation{Departamento de F\'{\i}sica, Universidade Federal do Amazonas, 67.077-000,
Manaus, Amazonas, Brazil}
\author{C. F. Destefani}
\affiliation{Department of Physics and Astronomy, Ohio University, 45701-2979, Athens,
Ohio, USA}
\author{Sergio E. Ulloa}
\affiliation{Department of Physics and Astronomy, Ohio University, 45701-2979, Athens,
Ohio, USA}
\date{\today }

\begin{abstract}
We have studied spin carrier dynamics under full spin-orbit coupling. The
anisotropy of dispersions for independent circular spinor polarizations is
explored as a possible vertical multichannel voltage controlled spin-filter.
Small voltage variations are found to select the current polarizations in a
resonant tunneling geometry.
\end{abstract}

\maketitle

The advent of \textit{spintronics} has resulted in the study and
design of spin manipulated devices used as information processors,
quantum computing elements, spin-polarized diodes, spin-valve read
heads,\cite{ausw} and electro-optical modulators,\cite{das} to
name just a few. The spin-orbit (SO) interaction, due to spatial
asymmetry in zincblende lattices or sample design, doping profile
or applied gate voltages, plays a fundamental role on these
systems,\cite{gem1,liu} especially on spin-dependent tunnelling.
The SO interaction couples the electronic momentum to the spin
degrees of freedom, and lifts the spin degeneracy for structures
fabricated in zincblende materials. The narrower the energy gap
of the host material, the stronger these effects will appear on the transport%
\cite{nitta} and optical properties.\cite{gem2}

We present here a detailed analysis of the complexity of bulk electronic
structure, derived from kinetic energy plus the \textit{full} SO Hamiltonian
including Bychkov-Rashba\cite{bych} and \textit{all three }Dresselhaus\cite%
{dress} contributions. Then, we study SO effects on spin-polarized current
in double-barrier resonant (DBR) devices, and analyze how the anisotropy on
the spin orientation and polarization of spinor states can be used to
\textit{select} and optimize channels for vertical transport in the system.

Recent publications have proposed the manipulation of linear Rashba and
Dresselhaus terms in the SO interaction.\cite{car} The cancellation of
\textit{linear} terms leads to drift-diffusive lateral transistors, in
contrast to the ballistic operation of the Datta-Das device.\cite{das} Perel
\textit{et al}\cite{perel} have also reported on the evolution of the spin
orientation with 2D transverse momentum, and explored the tunnelling through
a single barrier with different III-V materials, driven only by the linear $%
k_{\Vert }$ Dresselhaus term. A time-dependent spin manipulation scheme has
been recently proposed.\cite{re}

Let $\sigma =+$ ($-$) label the \textit{spin-up }(\textit{spin-down}) state.
The carrier dynamics, driven by the kinetic energy plus \textit{full} SO
Hamiltonian, is dictated by
\begin{equation}
\left[
\begin{array}{ll}
H_{_{++}} & H_{_{+-}} \\
H_{_{-+}} & H_{_{--}}%
\end{array}%
\right] \left[
\begin{array}{c}
F_{+}(z\mathbf{)} \\
F_{-}\mathbf{(}z\mathbf{)}%
\end{array}%
\right] e^{i\mathbf{k}_{\parallel }\mathbf{\cdot \rho }}=E\left[
\begin{array}{c}
F_{+}(z\mathbf{)} \\
F_{-}\mathbf{(}z\mathbf{)}%
\end{array}%
\right] e^{i\mathbf{k}_{\parallel }\mathbf{\cdot \rho }}\text{.}
\label{eq.1}
\end{equation}%
In this equation, the diagonal terms are $H_{\sigma \sigma }=-(\hbar
^{2}/2m^{\ast })d^{2}/dz^{2}+(\hbar ^{2}k_{\Vert }^{2}/2m^{\ast })-i\sigma
\alpha _{2}(k_{\Vert },\varphi )d/dz$, while the off-diagonal term is $%
H_{_{+\,-}}=\beta _{R}(k_{\Vert },\varphi )+\alpha _{3}(k_{\Vert },\varphi
)+\alpha _{1}(k_{\Vert },\varphi )d^{2}/dz^{2}$, with $H_{-+}=H_{+-}^{%
\dagger }$. In this context, $\mathbf{k}_{\parallel
}=(k_{x},k_{y})=(k_{\Vert },\varphi )$ is the in-plane wave vector, $\mathbf{%
\rho }=(x,y)$ is the electron position in the $xy$ plane, $\alpha
_{1}(k_{\Vert },\varphi )=\alpha _{0}k_{\Vert }e^{i\varphi }$, $\alpha
_{2}(k_{\Vert },\varphi )=\alpha _{0}k_{\Vert }^{2}\cos 2\varphi $ and $%
\alpha _{3}(k_{\Vert },\varphi )=\alpha _{0}k_{\Vert }^{3}(e^{i\varphi
}-e^{-i3\varphi })/4$ are, respectively, the linear, quadratic\ and cubic
Dresselhaus SO terms. Finally, $\beta _{R}(k_{\Vert },\varphi )=\beta
_{0}(-dV_{R}/dz)k_{\Vert }e^{i(\pi /2-\varphi )}$ is the linear
Bychkov-Rashba contribution under any potential profile $V_{R}(z)$, whereas $%
F_{\pm }(z\mathbf{)}$ are the spinor components. Here, $\alpha _{0}$ and $%
\beta _{0}$ are the SO material constants.\cite{param}

First let us discuss bulk dispersions where $k_{z}$ is a good quantum number
for the $-id/dz$ operator. Due to the strong coupling between spin $s$ and
linear momentum $\mathbf{k=(k}_{\Vert },k_{z})$ degrees of freedom, the
Dresselhaus (or bulk inversion asymmetry, BIA) and Bychkov-Rashba (or
surface inversion asymmetry, SIA) Hamiltonians create special sectors and
anisotropies on the spinor phase-space. For given $\mathbf{k}_{\Vert }$ and
applied electric field $F_{0}=-(1/e)dV_{R}/dz$, the bulk eigenvalues of the
full Hamiltonian in Eq.~\ref{eq.1} are
\begin{equation}
E_{\pm }(\mathbf{k}_{\Vert },k_{z})=\frac{\hbar ^{2}}{2m^{\ast }}(k_{\Vert
}^{2}+k_{z}^{2})\pm \sqrt{\gamma (\mathbf{k}_{\Vert },k_{z})+\delta (\mathbf{%
k}_{\Vert })}\text{,}  \label{eq.2}
\end{equation}%
where $\gamma (\mathbf{k}_{\Vert },k_{z})=\alpha _{0}^{2}k_{\Vert
}^{4}k_{z}^{2}\cos (4\varphi )+\alpha _{0}^{2}k_{\Vert
}^{2}k_{z}^{4}-2\alpha _{0}\beta k_{\Vert }^{2}k_{z}^{2}\sin (2\varphi )$
and $\delta (\mathbf{k}_{\Vert })=\beta ^{2}k_{\Vert }^{2}+\alpha
_{0}^{2}k_{\Vert }^{6}\sin ^{2}(2\varphi )/4+\alpha _{0}\beta k_{\Vert
}^{4}\sin (2\varphi )$. Here $\varphi $ is measured from the $(1,0)$
crystalline axis on the $xy$ plane and $\beta =\beta _{0}(-dV_{R}/dz)$.

Notice that from the four degeneracies, two for the momentum ($\pm k_{z}$)
and two for spin ($\pm $), one can construct only two independent spinor
states forming the Kramers doublet. Without SO, there are two degenerate
states, well known as \textit{linear spin polarized}, where the \textit{%
spin-up} and \textit{spin-down} spinors display isotropic parabolic
dispersions for both $k_{\Vert }$ and $k_{z}$ (first term in Eq.~\ref{eq.2}%
). However, with SO there are two orthogonal Hilbert sub-spaces, the \textit{%
circular spin polarized} states, with highly anisotropic dispersions when
the orbital angle, $\varphi $, changes clockwise ($\sigma ^{+}$) and
counterclockwise ($\sigma ^{-}$), respectively. The gap measuring the
spin-splitting is $E_{g}^{\sigma }(k_{\Vert },\varphi
,k_{z})=[E_{+}(k_{\Vert },\varphi ,k_{z})-E_{-}(k_{\Vert },\varphi ,k_{z})]$%
, or twice the square root in Eq.~\ref{eq.2}. At $k_{z}=0$, this yields a
critical angle $\varphi _{c}^{\sigma }=-0.5\sin ^{-1}[-2\beta /(\alpha
_{0}k_{\Vert }^{2})]$\ where $E_{g}^{\sigma }(k_{\Vert },\varphi
_{c}^{\sigma },k_{z})=0$. This result depends on the field strength and on
the ratio between SO constants, and defines a maximum value for $k_{\Vert }$%
, given by $k_{\Vert }^{c}=\sqrt{-2\beta /\alpha _{0}}$. There are two
symmetry regimes according to which SO mechanism is dominant: (a) \textit{%
SIA regime}, or Rashba dominant, for $k_{\Vert }\leq k_{\Vert }^{c}$, and
(b) \textit{BIA regime}, or Dresselhaus dominant, for $k_{\Vert }>k_{\Vert
}^{c}$. These properties are valid for any zincblende material and here we
will fix our attention on InSb, due to its large SO parameters.\cite{param}
These states and their properties are reminiscent of circularly-polarized
light travelling through a crystal, as we shall discuss later. It is
convenient to define a new zero for the angle at the $(1,-1)$ crystalline
direction in the $xy$ plane, corresponding to a rotation $\varphi
\Rightarrow \varphi -45%
{{}^\circ}%
$. We adopt this from now on.

\begin{figure}
  \includegraphics[width=8.5cm]{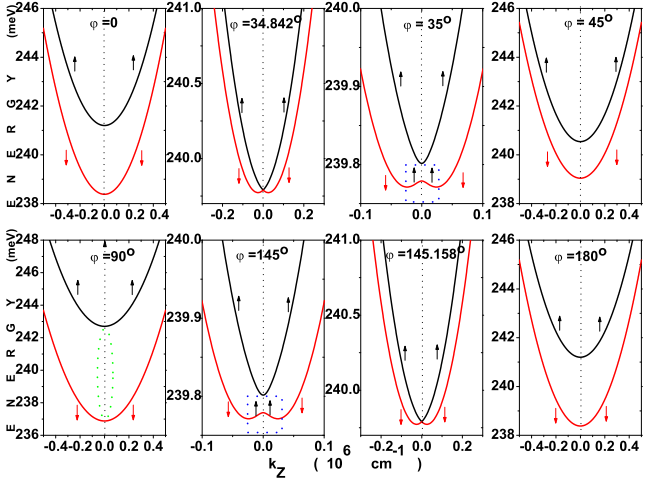}\\
  \caption{Angular anisotropies and bulk dispersions for $\protect\sigma ^{+}$%
-polarization at special angles within the interval $0\leq \protect\varphi %
\leq 180%
{{}^\circ}%
$. Notice spin polarization in different sectors of $k_{z}$.
Panels
at $\protect\varphi =\protect\varphi _{c}^{\protect\sigma ^{+}}=34.842%
{{}^\circ}%
$ and at $\protect\varphi =145.158%
{{}^\circ}%
$ show no spin gap.}
  \label{fig1}
\end{figure}

Figure \ref{fig1} shows different cuts on the energy surface, for the BIA
regime and $\sigma ^{+}$ polarization, calculated for a fixed value of $%
k_{\Vert }=3$, and different values of $\varphi $ when a uniform field, $%
F_{0}=5$, is applied. For InSb\cite{param} and the given $F_{0}$, the
critical values are $\varphi _{c}^{\sigma ^{+}}=34.842%
{{}^\circ}%
$ and $k_{\Vert }^{c}=1.768$. The panel for $\varphi =0$ shows the
non-parabolic spin-splitted dispersions for \textit{spin-up} and \textit{%
spin-down} branches. For a given energy inside the gap region, the \textit{%
spin-up} states have imaginary values for $k_{z}$ along a \textit{real line}%
, shown as dotted green lines in the panel with $\varphi =90%
{{}^\circ}%
$, and these states do not propagate. The next panel with $\varphi =\varphi
_{c}^{\sigma ^{+}}$ shows the situation where $E_{g}(k_{\Vert },\varphi
_{c}^{\sigma ^{+}},k_{z}=0)=0$. Just above this angle, at $\varphi =35%
{{}^\circ}%
$, the gap opens again and an \textit{inverted} group-velocity layer of
\textit{spin-up} states (inside dotted blue-box) appears between critical $%
k_{z}$-values, $\pm k_{c1}$, where the $z$-group-velocity vanishes. The
existence of this branch is determined partly by the quadratic $\alpha
_{2}(k_{\Vert },\varphi )$ BIA term in Eq.~\ref{eq.1}. For $k_{z}$ in the
vicinity of these critical points we have $E_{+}(\mathbf{k}_{\Vert
},k_{z})\sim E_{0}+S_{0}(k_{z}\pm k_{c1})^{2}$, with $E_{0}$ and $S_{0}$
constants. Inside the dotted blue box the state has \textit{inverted}
group-velocity and negative effective mass, such that a carrier will travel
to $\mp z$-axis for $\pm k_{z}$ values of the propagating wave vector. In
the higher \textit{spin-up} and \textit{spin-down} branches outside $\pm
k_{c1}$,\ the carriers travel with normal group-velocity. By increasing the
angle to $\varphi =45%
{{}^\circ}%
$, the double-valley region disappears, with an increasing gap that reaches
its maximum value at $\varphi =90%
{{}^\circ}%
$. The panels with $90%
{{}^\circ}%
\leq $ $\varphi $ $\leq 180%
{{}^\circ}%
$ just complete half of the periodicity of the $\sigma ^{+}$ Hilbert
subspace.

\begin{figure}
  \includegraphics[width=8.5cm]{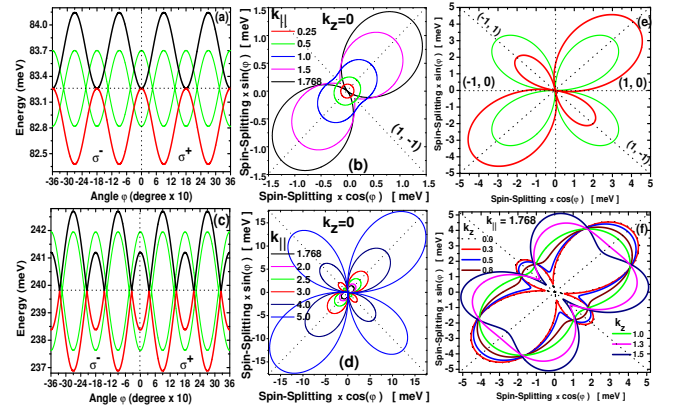}\\
  \caption{Spin-splitting
anisotropy for \textit{SIA regime} (at $k_{\Vert }=1.768=k_{\Vert
}^{c}$,
panel $a$) and \textit{BIA regime} (at $k_{\Vert }=3>k_{\Vert }^{c}$, panel $%
c$) produced by full SO (black/red lines) and BIA terms only
(green lines) in bulk InSb. Panels $b$ and $d$: Twofold and
fourfold symmetries for
spin-splitting gap in the \textit{SIA} ($b$) and \textit{BIA} ($d$) \textit{%
regimes}, at different values of $k_{\Vert }$. Panel $e$:
Comparison between Rashba and Dresselhaus interactions, for panel
$c$ conditions; red (green)
lines for BIA+SIA\ (BIA) terms. Panel $f$: Spin-splitting anisotropy of $%
\protect\sigma ^{+}$ energy surfaces, away from zone-center.}
  \label{fig2}
\end{figure}

The dispersions for $\sigma ^{-}$ polarization, in this BIA regime, have
identical features as in Fig.~\ref{fig1}, except that they require exchange
between spin-polarizations in the sectors. The Hilbert subspaces form the
degenerate Kramers doublet for circular polarizations. These $\sigma ^{-}$
and $\sigma ^{+}$ spinors are characterized by their behavior under the
time-reversal operator for zincblende symmetry, $\widehat{\Upsilon }=-i%
\widehat{\sigma }_{y}\widehat{\complement }\widehat{I}$, where $-i\widehat{%
\sigma }_{y}$ flips spin and the complex conjugation (inversion) $\widehat{%
\complement }$ ($\widehat{I}$) flips momentum (position).

Figure \ref{fig2}a shows the angular gap variation, for both circular
polarizations, at the critical situation where $k_{\Vert }=k_{\Vert
}^{c}=1.768$, for $F_{0}=5$. Black and red lines are the combined results
for BIA+SIA terms whereas green lines show the effect of BIA terms acting
alone. It becomes clear that BIA+SIA terms act in-phase and out-of-phase in
each sector. This alternating change in phase between the SO Hamiltonian
terms leads to a \textit{Rashba regime} with $C_{2V}$ symmetry, as shown in
Fig.~\ref{fig2}b for different $k_{\Vert }\leq k_{\Vert }^{c}$. In this
regime, the Hamiltonian in Eq.~\ref{eq.1} produces a smooth change from $s$-
to $p$-like symmetry on the electronic structure at the zone-center as $%
k_{\Vert }$ increases.

Figure~\ref{fig2}c shows the angular gap variation in the \textit{%
Dresselhauss regime} (black and red lines for $k_{\Vert }=3>k_{\Vert }^{c}$,
for $F_{0}=5$). The effect of BIA terms alone is shown in green lines. Here
also the full SO Hamiltonian acts in-phase and the overall symmetry is
changed from a twofold $C_{2V}$ to a fourfold $C_{4V}$. As shown in Fig.~\ref%
{fig2}d, this modification induces smooth changes on the electronic
structure at the zone-center, from $p$- to $d$-like symmetry as $k_{\Vert }$
increases. This is also at the origin of the secondary gap and the
double-valley sector near the critical angle. The resulting symmetries show
that it is essential to consider all the three BIA terms in the Hamiltonian,
so that their cumulative effects yield the correct electronic structure and
eigenstates. Observe that at given electric field there are only specific
angles where the Rashba and Dresselhaus effects cancel each other.

\begin{figure}
  \includegraphics[width=8.5cm]{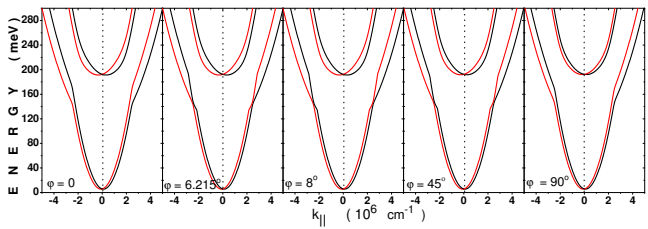}\\
  \caption{2D-subband dispersions, solutions of
Eq.~\protect\ref{eq.1}, in isolated InSb QW with $L_{w}=200$ \AA
\thinspace\ and uniform electric field $F_{0}=100$. Different
anisotropies induced by
the full SO terms result in varying in-plane masses, $m^{\ast }(k_{\Vert },%
\protect\varphi )$, and spin-splittings for each subband. Red
(black) lines for down (up) spin.}
  \label{fig3}
\end{figure}

Figure \ref{fig2}e compares the combined effects of BIA and SIA terms (red
lines) on the spin-splitting, at the zone-center, for the same conditions
shown in Fig.~\ref{fig2}c and the red line in Fig.~\ref{fig2}d (notice scale
changing). It is clear that their effects are added along the $(\pm 1,\pm 1)$
directions of maximum spin-splitting and the BIA terms (green lines)
dominate along $(\pm 1,\mp 1)$ directions where the secondary gap appears.

If we move away from the zone-center, the complexity increases. In Fig.~\ref%
{fig2}f, we start with the critical situation displayed in Figs.~\ref{fig2}a
and \ref{fig2}b. The cuts at increasing values of $k_{z}$ show new gap
maxima along the $(\pm 1,0)$ and $(0,\pm 1)$ directions. These curves are a
reflection of the complex geometry of the 2D Fermi surface at different
electron concentrations.

Evidence of strong anisotropy on FIR light propagating through n-type InSb
bulk samples was studied by Gopalan \textit{et al}.\cite{gopa} They have
shown why the magnetotransmission experiments of Dobrowoslka \textit{et al},%
\cite{dobro} in the Voigt and Faraday configurations, displayed maximum
spin-flip and maximum transmissivity along the $(\pm 1,\pm 1)$ and $(\pm
1,\mp 1)$ directions. The electronic structure we report here arises from
the same complex lattice symmetry.

Having shown the complexity of the bulk spinor environment under the full SO
Hamiltonian, let us now explore these anisotropies via the vertical
transport of carriers in a DBR structure with interfaces [thickness] at
positions $z_{\ell }$ $(\ell =1,...,4)$ [$d_{l}=(z_{\ell +1}-z_{\ell })$]
and profile $V_{0}(z)$. The layer $\ell =0$ ($\ell =4$) is the emitter
(collector) contact. We use the scattering matrix technique to calculate the
transmissivity and reflectivity for the system. For a given incident energy,
$E$, we want to construct the \textit{incoming} and \textit{outgoing} spinor
states that tunnel through the 2D states in the layer $\ell =2$. In Fig.~\ref%
{fig3} we show 2D dispersions derived from Eq.~\ref{eq.1}, in an isolated
quantum well under applied electric field. We observe that the 3D
anisotropies are transferred to the subband dispersions, through anisotropic
in-plane effective masses $m^{\ast }(k_{\Vert },\varphi )$, as well as
spin-splittings of each subband.

We construct each component of a spinor state, $F(\rho ,\mathbf{k}_{\Vert
},z)$ in Eq.~\ref{eq.1}, as a linear combination of bulk solutions\cite{bit}
$\Phi (z)$ = $\sum_{\sigma (+,-)}[a_{l}^{\sigma }(\mathbf{k}_{\Vert
})F_{\sigma }(+k_{z\sigma })e^{+ik_{z\sigma }(z-z_{l})}$ + $b_{l}^{\sigma }({%
\mathbf{k}_{\Vert }})F_{\sigma }(-k_{z\sigma })e^{-ik_{z\sigma }(z-z_{l})}]$%
, where $a_{l}^{\sigma }(\mathbf{k}_{\Vert })$ ($b_{l}^{\sigma }(\mathbf{k}%
_{\Vert })$) is the amplitude of \textit{incoming} (\textit{outgoing})
spin-polarized waves at any given interface $z_{l}$, and $\pm k_{z\sigma }$
are real roots of $E_{\sigma }(\mathbf{k}_{\Vert },k_{z})=E$, for the
carriers travelling along the $z$-axis. The dependence on $\mathbf{k}%
_{\parallel }$ defines the open channels in phase-space. The current
operator from Eq.~\ref{eq.1} is
\begin{equation}
J_{z}(\mathbf{\mathbf{k}_{\Vert }},k_{z})=\left[
\begin{array}{cc}
J_{++} & J_{+-} \\
J_{-+} & J_{--}%
\end{array}%
\right] \text{,}  \label{eq.3}
\end{equation}%
where $J_{\sigma \sigma }=\hbar k_{z}/m^{\ast }+\sigma \alpha _{2}(k_{\Vert
},\varphi )/\hbar $ represents the spin-conserving polarized current, and $%
J_{+-}=-2m^{\ast }\alpha _{1}(k_{\Vert },\varphi )k_{z}/\hbar $ the
spin-flip mechanism. The vertical transport properties are also determined
by the quadratic (diagonal) and linear (spin-flip) BIA SO terms. The
boundary conditions require that both spinor $F(\rho ,\mathbf{k}_{\Vert
},z=z_{l})$ and flux $J_{z}F(\rho ,\mathbf{k_{\Vert },}z=z_{l})$ be
continuous across each interface, $z_{l}$. Resonant tunnelling for a given
circular polarization implies that the matching conditions must be valid for
each $\mathbf{k}_{\Vert }$.

\begin{figure}
  \includegraphics[width=8.0cm]{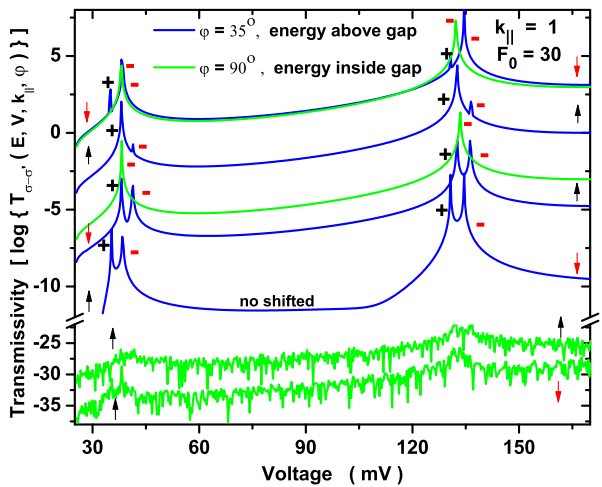}\\
  \caption{Transmissivity for $\protect\sigma ^{+}$ spinor in InSb-InAlSb double
barrier with $l_{b}=100$ \AA\ and $L_{w}=200$ \AA , calculated for
$k_{\Vert
}=1$, $F_{0}=30$ and $\protect\varphi =35%
{{}^\circ}%
$ ($90%
{{}^\circ}%
$) for an incident energy above (inside) the gap, shown in blue
(green) lines. The up/down arrows on left (right) indicate
incidence (detection) polarizations in the DBR. Curves are shifted
upwards for clarity, and small differences in peak positions of
the two highest curves are due to distinct values of
$\protect\varphi $. Notice different amplitudes of each resonant
peak at spin-polarized 2D levels inside QW. For energies inside
the gap, only one spin propagates.}
  \label{fig4}
\end{figure}

The elements of the scattering matrix $\mathbf{S}$ are the spin-conserving ($%
t_{+}^{+}$ and $t_{-}^{-}$) and spin-flip ($t_{-}^{+}$ and $t_{+}^{-}$)
transmission, and the equivalent reflection coefficients $r_{\pm }^{\pm }$,
associated to the \textit{incoming} and \textit{outgoing} wave amplitudes $%
t_{\sigma ^{\prime }}^{\sigma }=a_{4}^{\sigma ^{\prime }}/a_{0}^{\sigma }$,
for detected ($\sigma ^{\prime }$) and incident ($\sigma $) polarizations.
For any applied potential $V$, the partial transmissivity can be calculated%
\cite{bit} from the ratio between detected $J_{\sigma ^{\prime }}^{out}$ and
incident $J_{\sigma }^{in}$ currents as $T_{\sigma \rightarrow \sigma
^{\prime }}(E,V,\mathbf{k}_{\Vert })=\Re e[(t_{\sigma ^{\prime }}^{\sigma
})^{\ast }t_{\sigma ^{\prime }}^{\sigma }\langle F_{\sigma ^{\prime
}}^{+}|J_{z\sigma ^{\prime }}^{+}|F_{\sigma ^{\prime }}^{+}\rangle
_{z=z_{4}}/$ $\langle F_{\sigma }^{+}|J_{z\sigma }^{+}|F_{\sigma
}^{+}\rangle _{z=z_{0}}]$. With the presence of spin-flip processes, the
partial reflectivity is calculated from the general flux conservation, $%
\sum_{\sigma ^{\prime }}[T_{\sigma \rightarrow \sigma ^{\prime
}}(E,V)+R_{\sigma \rightarrow \sigma ^{\prime }}(E,V)]=1$, for each incident
polarization. We consider vertical transport along the $z$-direction $(001)$
of a symmetric InSb-InAlSb\cite{param} DBR [$l_{b}=100$ \AA\ (barrier) and $%
L_{w}=200$ \AA\ (well)].

Figure \ref{fig4} shows calculated partial transmissivities in the $\sigma
^{+}$ polarization. In principle, at a given incident energy $E$, there will
be transmissivity peaks at each spin-polarized QW level. As mentioned above,
\textit{spin-up} states cannot propagate when $E$ is in the gap-region since
the $k_{z\sigma }$ roots of $E_{\sigma }(\mathbf{\mathbf{k}_{\Vert },}%
k_{z})=E$ are imaginary, as shown in panel $\varphi =90%
{{}^\circ}%
$ of Fig.~\ref{fig1}. For $E$ outside the gap-region all polarizations
propagate. The four transmissivity curves in Fig.~\ref{fig4} for an energy
above the gap (blue lines) show double peaks at the quantum well resonances,
for $\sigma ^{+}$-polarization. These same features occur for $\sigma ^{-}$%
-polarization at equivalent angles. However, when the incident energy is
inside the gap (green lines), notice that only the \textit{spin-down}
incident polarization can propagate, while the \textit{spin-up} channels
result in vanishing transmissivity. The resulting current arriving at the
collector will have Stark shifted double peaks in voltage for each quadrant.%
\cite{bit} Since $\sigma ^{+}$- and $\sigma ^{-}$-polarizations belong to
orthogonal Hilbert subspaces, the vertical current produced by these states
(for energies above the gap) define eight independent emitter-collector
transmission channels that may be explored as voltage controlled
spin-filters. In fact, polarized currents with desired polarization ($\pm $)
can be produced with small voltage variations. Notice also that for energies
inside the gap the total current is, \textit{de facto,} \textit{spin-down}
polarized.

In conclusion we have studied how anisotropies of spinor states and
dispersions of the \textit{full} SO Hamiltonian in orthogonal Hilbert
spaces, with opposite circular spin-polarized configurations, could enhance
the formation of spin-filters. The possibility to tune emitter Fermi level
and collector detectors requires designed magnetic contact masks on the DBR
diode, a technology in full development nowadays.\cite{ras2} This would help
exhibit and explore the complexity of the eight channels produced by the
\textit{full} SO Hamiltonian. We have shown that this new structure can work
as multichannel voltage controlled spin-filters, similar to semimagnetic
materials.\cite{molek} We anticipate that new designed experiments will soon
be able to probe the gap structure and anisotropies we predict. Finally, the
possible scattering mechanisms within these two orthogonal spinor
configurations must be analyzed under the unusual situation where spin and
momentum coordinates are strongly coupled. Therefore, new carrier \textit{%
spin-flip relaxation} times, $\tau _{SF}$, proportional to the reciprocal of
spin-splitting energy,\cite{dyako} $E_{g}^{\sigma }(k_{\Vert },\varphi
_{c},k_{z})$, and momentum scattering times, $\tau _{p}$, can be measured
since $\tau _{SF}^{-1}=\tau _{p}\left[ \frac{E_{g}^{\sigma }(k_{\Vert
},\varphi _{c},k_{z})}{\hbar }\right] ^{2}/2$. The very large values for $%
E_{g}^{\sigma }(k_{\Vert },\varphi _{c},k_{z})$ ($\sim 5-10$ meV) can
produce devices with $\tau _{SF}$ many orders faster than $\tau _{p}$.

Work is supported by FAPESP, CNPq, CAPES, US DOE grant no.\
DE-FG02-91ER45334, and the CMSS Program at OU.

\end{document}